\documentclass[pra,twocolumn,showpacs,floatfix]{revtex4}
\usepackage{graphicx}
\begin{document}

\title{Mixtures of Bose gases confined in a ring potential}
\author{J. Smyrnakis$^1$, S. Bargi$^2$, G. M. Kavoulakis$^1$, 
M. Magiropoulos$^1$, K. K\"arkk\"ainen$^2$, and 
S. M. Reimann$^2$}
\affiliation{$^1$Technological Education Institute of Crete, 
P.O. Box 1939, GR-71004, Heraklion, Greece \\
$^2$Mathematical Physics, Lund Institute of Technology, 
P.O. Box 118, SE-22100 Lund, Sweden}
\date{\today}

\begin{abstract}

The rotational properties of a mixture of two 
distinguishable Bose gases that are confined in a ring 
potential provide novel physical effects that we 
demonstrate in this study. Persistent currents are shown 
to be stable for a range of the population imbalance between 
the two components at low angular momentum. At higher values 
of the angular momentum, even small admixtures of a second 
species of atoms make the persistent currents highly fragile.

\end{abstract}
 \maketitle

{\it Introduction.}
One of the most fascinating phenomena associated 
with superfluidity \cite{Leggett} is the stability 
of persistent currents. In some remarkable experiments 
that have been performed recently, Bose-Einstein 
condensed atoms were confined in annular traps 
\cite{Kurn,Olson}, in which persistent currents 
could be created and observed \cite{Phillips}. In 
an earlier experiment, the resistant-free motion 
of an object through a Bose-Einstein condensate 
below some critical velocity, was also observed 
\cite{Ketterle}.

Motivated by these recent advances, in the present
study we consider a mixture of two (distinguishable) 
Bose gases at zero temperature \cite{Erich,Eric}, 
that are confined to one dimension with periodic 
boundary conditions, i.e. in a ring potential, deriving 
a series of exact and analytic results.

The main issue of our study concerns the rotational 
properties of this system and the stability of persistent 
currents. In higher dimensions it has been argued that 
mixtures of Bose gases do not support persistent currents, 
because there is no energy cost for the system to get rid 
of its circulation (i.e., the line integral of the velocity
field around a closed loop that encircles the ring), 
as long as angular momentum can be transferred between 
the two species \cite{Ho}. Here, we demonstrate that when 
the total angular momentum per atom varies between zero and 
unity, currents are stable for a certain range of the ratio 
of the populations of the two species. We calculate the 
critical strength of the coupling for a given value of 
this ratio, which we determine analytically and exactly. 
On the other hand, for higher values of the angular 
momentum per atom, persistent currents in one-component 
systems are very fragile, as even small admixtures 
of a second species of atoms destabilize the currents.

{\it Model.}
Assuming a ring potential (which corresponds to a 
very tight annular trap along the transverse direction
\cite{1D}), the Hamiltonian of the system that we study 
for the two components that we label as $A$ and $B$ is 
$H = H_{AA} + H_{BB} + {\tilde U}_{AB} 
\sum_{i=1, j=1}^{N_A, N_B} \delta(\theta_{i} - 
\theta_{j})$, where
\begin{eqnarray}
   H_{kk} = \sum_{i=1}^{N_k} - \frac {\hbar^2} {2 M_k R^2} 
 \frac {\partial^2} {\partial \theta_{i}^2} 
+ \frac 1 2 {\tilde U}_{kk} \sum_{i \neq j=1}^{N_k} 
\delta(\theta_{i} - \theta_{j}),
\end{eqnarray}
with $k=A,B$. Here $M_k$ are the atom masses, while
${\tilde U}_{kk} = 4 \pi \hbar^2 a_{kk}/(M_{k} R S)$
and ${\tilde U}_{AB} = 2 \pi \hbar^2 a_{AB}/(M_{AB} R S)$ 
are the matrix elements for zero-energy elastic atom-atom 
collisions (all assumed to be positive), with $M_{AB}=M_A 
M_B/(M_A + M_B)$ being the reduced mass. Also, $R$ is the 
radius of the annulus and $S$ its cross section, with 
$R \gg \sqrt S$.

We start from the mean-field approximation, introducing 
the order parameters of the two components ${\phi}_A$ and 
${\phi}_B$; later we also go beyond the mean-field 
approximation, diagonalizing the Hamiltonian $H$
numerically and analytically. The resulting (coupled) 
nonlinear Gross-Pitaevskii-like equations are
\begin{eqnarray}
  - \frac {\partial^2 {\phi}_k}{\partial \theta^2}
+ N_k U_{kk} |{\phi}_k|^2 {\phi}_k + N_l U_{kl} |{\phi}_l|^2 
{\phi}_k &=& {\mu}_k {\phi}_k,
\label{gpe}
\end{eqnarray}
where $\int |{\phi}_k|^2 d \theta = 1$. Here $\mu_k$ are 
the chemical potentials divided by the kinetic energy 
$\varepsilon = \hbar^2/(2MR^2)$, where we have assumed for 
simplicity equal masses for the two species, $M_A = M_B 
= M$. Also, $U_{kl} = {\tilde U}_{kl}/\epsilon$, with 
$k,l=A,B$.

{\it Energetic stability, dynamic stability and phase 
separation.}
Before we turn to the rotational properties, let 
us consider briefly the question of phase separation.
In homogeneous systems it has been shown that the 
condition for energetic stability of the homogeneous 
solution is \cite{Ao,Timm,Chris} $U_{AB}^2 - U_{AA} 
U_{BB} < 0$, and also $U_{AA}>0$, $U_{BB}>0$. One may 
generalize this result for the case of a finite system, 
taking into account the contribution of the kinetic 
energy. The details of this calculation will be reported 
elsewhere. Here we just mention that this more
general condition is 
$\gamma_{AB}^2 - \gamma_{AA} \gamma_{BB} < 1/4 + 
(\gamma_{AA} + \gamma_{BB})/2$, where we have 
introduced the parameters $\gamma_{k,l} = U_{k,l} 
\sqrt{N_k N_l}/(2 \pi)$ for convenience (these 
parameters give the ratio between the typical 
interaction energy and the typical kinetic energy). 
As one crosses the phase boundary, the two clouds 
develop sinusoidal variations in their density, 
with an amplitude that increases continuously from 
zero.

The dynamic stability of the system may be examined
with use of the (two coupled) Bogoliubov-de Gennes 
equations. Again, the details of this calculation
will be reported elsewhere. The dispersion that one
obtains from this analysis is $\omega^2 = m^4 + m^2 
\left( \gamma_{AA} + \gamma_{BB} \pm \sqrt{(\gamma_{AA} 
- \gamma_{BB})^2 + 4 \gamma_{AB}^2} \right)$.
The requirement of a real $\omega$ implies the same 
condition as that for energetic stability. 

{\it Effect of the periodicity on the dispersion relation.} 
The one-dimensional motion that we have assumed in 
our calculation, in combination with the periodic 
boundary conditions have some important consequences 
on the dispersion relation, which are also present 
in the case of a single-component gas, as shown by 
Bloch \cite{Bloch}. The matrix elements that determine 
the interaction energy do not depend on the quantum 
numbers of the angular momentum $m$, and also the 
center of mass coordinate separates from the relative 
coordinates. As a result, solving the problem in the 
interval $0\le l \le 1$, where $l = (L_A + L_B)/
(N_A + N_B)$ is the angular momentum per particle, 
then exciting the center of mass motion, we may 
evaluate the spectrum at any other interval $n \le 
l \le n+1$. More specifically, if $\phi_{A,0} = 
\sum_m c_m \Phi_m$, and $\phi_{B,0} = \sum_m d_m 
\Phi_m$, are the order parameters for $0\le l \le 1$, 
then the order parameters for $n \le l \le n+1$ are
given by ${\phi}_{A,n} = \sum_m c_m \Phi_{m+n}$, 
and ${\phi}_{B,n} = \sum_m d_m \Phi_{m+n}$. 

Denoting the energy per atom for $n \le l \le n+1$ as 
$E_n(l)/N$, then $E_n(l)/N = E_0(l_0)/N + n^2 + 2 n l_0$, 
where $0 \le l_0 \le 1$, and $l = l_0 + n$. Therefore, 
$E_n(l)/N - l^2 = E_0(l_0)/N - l_0^2$, which are both 
equal to a periodic function $e(l)$, i.e., $e(l_0 + n) 
= e(l_0)$. Thus, we write quite generally that
\begin{eqnarray}
  E_n(l)/N = l^2 + e(l) = (l_0+n)^2 + e(l_0).
\label{Bl}
\end{eqnarray}
In other words, the energy of the system for $n \le l 
\le n+1$ consists of an envelope part, i.e., the first 
term on the right, which arises because of the center 
of mass excitation, plus a periodic part $e(l)$. 

Furthermore, the function $e(l_0)$ is symmetric around 
$l_0=1/2$ (an example of this symmetry is demonstrated 
below, where it is shown that $E_0/N$ is linear for 
$0 \le l \le x_B = 1-x_A$ and $x_A \le l \le 1$). To see 
this, let us consider the states ${\phi}_{A}^R = \sum_m 
c_m \Phi_{1-m}$, and ${\phi}_{B}^R = \sum_m d_m \Phi_{1-m}$,
with an $l'$ equal to $1-l$, or $l+l'=1$. It turns out 
that the difference in the energy per particle in the 
states ${\phi}_{A}^R$, ${\phi}_{B}^R$, and ${\phi}_{A}$,
${\phi}_{B}$ is $\Delta E/N = l' - l$. However,
according to Eq.\,(\ref{Bl}), $\Delta E/N = l'-l+e(l')
-e(l)$, and therefore $e(l')=e(l)$, which means that 
$e(l_0)$ is indeed symmetric around $l_0=1/2$.

{\it Rotational properties.} Since, according to what
was mentioned above, the dispersion relation is 
quasi-periodic, in order to study the rotational 
properties of the gas, we restrict ourselves to the 
interval $0 \le l \le 1$. We introduce the variables 
$x_A = N_A/(N_A + N_B)$ and $x_B = N_B/(N_A + N_B)$, 
and assume without loss of generality that $x_B < x_A$, 
with $x_A + x_B = 1$. In what follows we also assume 
equal scattering lengths, and therefore $U_{AA} = U_{BB} 
= U_{AB} = U$. The condition of equal scattering 
lengths is not far from reality, with Rubidium atoms 
in different hyperfine states being an example. 
Interestingly, in this case there is a series of 
exact, analytic results. If this condition is 
weakly violated, the deviations from these results 
will be small. 

According to the result mentioned earlier, for 
$U_{AA} = U_{BB} = U_{AB}$ the gas is in the 
homogeneous phase, and it is both dynamically, 
as well as energetically stable. In this case, 
we find that for $0 \le l \le x_B$ and $x_A \le 
l \le 1$, only the states with $\Phi_0$ and 
$\Phi_1$ are (macroscopically) occupied. The 
interaction energy of the gas is equal to that 
of the non-rotating system, since the total 
density $n(\theta) = n_A(\theta) + n_B(\theta)$ 
is homogeneous. As a result, the total energy of 
the gas varies linearly with $l$. These are exact 
results within the mean-field approximation. On 
the other hand, for $x_B < l < x_A$ more states 
contribute to the order parameters, while the 
dispersion relation is not linear in this interval. 
More specifically, let us consider the states of 
some fixed expectation value of the angular momentum 
$l$, ${\phi}_{A,0} = c_0 \Phi_0 + c_1 \Phi_1$, and 
${\phi}_{B,0} = d_0 \Phi_0 + d_1 \Phi_1$, with 
$x_A |c_1|^2 + x_B |d_1|^2 = l$, and also $|c_0|^2 
+ |c_1|^2 = 1$, $|d_0|^2 + |d_1|^2 = 1$. The above 
states have a maximum value of $l$ equal to unity. 
Evaluating the total energy $E_0$ and minimizing it, 
it turns out that 
\begin{eqnarray}
  E_0/N = l
+ {\gamma} \left[  1/2 + (x_A |c_0| |c_1| - 
x_B |d_0| |d_1|)^2 \right],
\label{mfen}
\end{eqnarray}
where $N = N_A + N_B$ is the total number of atoms
and $\gamma = N U/(2 \pi)$. For $0 \le l \le x_B$ 
and $x_A \le l \le 1$, the last two terms may be set 
equal to each other, which means that $E_0/N = l + 
\gamma/2$. Remarkably, any other single-particle 
state cannot lower the energy and its occupancy is 
exactly zero. The occupancies of the single-particle 
states with $m=0$ and $m=1$ are $c_0^2 = (x_A - l) 
(1 - l) /[x_A (1 - 2l)]$, and $c_1^2 = l (x_B - l)
/[x_A (1 - 2l)]$; $d_0^2$ and $d_1^2$ are given by 
similar formulae, with $x_A$ and $x_B$ interchanged. 
The same expressions hold for a mixtures of two Bose 
gases that are confined in harmonic traps \cite{BCKR}, 
but in this case the energy is parabolic and not linear 
in $l$.

{\it Persistent currents.}
Let us now examine the question of stability of persistent 
currents. In the case of only one component, for $\gamma 
> 3/2$, the system supports persistent currents at $l=1$ 
\cite{KSU,GMK}. As we saw earlier, if one starts with $x_A 
= 1$ and $x_B = 0$ and increases the population of the $B$ 
component, the dispersion relation is exactly linear for 
$x_A \le l \le 1$. The question is thus whether the
dispersion relation has a local minimum at $l=x_A$, 
where we know the order parameters exactly, i.e., 
$\phi_{A,0} = \Phi_1$, and $\phi_{B,0} = \Phi_0$. This 
fact allows us to examine the region just below $l = x_A$ 
(and the region just above $l = x_B$, if necessary). 

More specifically, if $\epsilon = x_A - l$ is a small and 
positive quantity, one may argue that $c_0^2 \propto c_2^2 
\propto \epsilon$, while $d_{-1}^2 \propto d_{1}^2 \propto 
\epsilon$. The asymmetry between the two species arises 
because $c_1 = 1$ and $d_0 = 1$ at $l=x_A$. As a result, 
for component $A$, $c_0 c_1^2 c_2 \propto c_1^2 c_2^2$, 
which implies that $c_2 \propto c_0$, while for component 
$B$, $d_{-1} d_0^2 d_1 \propto d_{-1}^2 d_0^2$, and 
thus $d_{-1} \propto d_1$. All the other coefficients 
are of higher order in $\epsilon$, and thus negligible 
as $l \to x_A^-$. Since the stability of the persistent 
currents is determined from the slope of the dispersion 
relation, we may keep only the terms which are linear in 
$\epsilon$. 
\begin{figure}
\includegraphics[width=8.5cm,height=5.5cm]{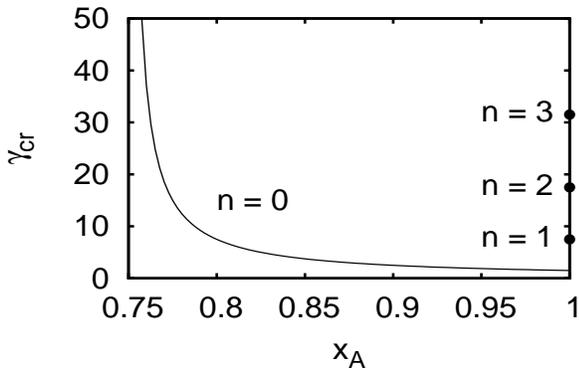}
\caption{The critical coupling $\gamma_{\rm cr}$ given
by Eq.\,(\ref{dcr}), in the interval $0 \le l \le 1$, 
as a function of $x_A$, for a ring potential. The points
at $x_A = 1$ show $\gamma_{\rm cr}$ for the higher 
intervals of $l$, as explained in the text.}
\label{FIG1}
\end{figure}
Under these assumptions we find that the energy per particle 
is, up to $\epsilon$,
\begin{eqnarray}
  E_0/N - \gamma/2 \approx l 
+ 2 x_A c_2^2 + 2 x_B d_{-1}^2 +
\nonumber \\
\gamma \left[ x_A (c_0+c_2) + x_B (d_{-1}+d_1) \right]^2,
\end{eqnarray} 
where we have expressed $c_1$ in terms of $c_0$ and $c_2$, 
and $d_0$ in terms of $d_{-1}$ and $d_1$ through the 
normalization conditions. The above expression has to be 
minimized under the constraint of fixed angular momentum, 
$l = x_A (c_1^2 + 2 c_2^2) + x_B (-d_{-1}^2 + d_1^2) = x_A 
- \epsilon$. We do this by minimizing the function $E_0/N
+ \lambda [x_A (c_1^2 + 2 c_2^2) + x_B (-d_{-1}^2 + d_1^2)]$, 
where $\lambda$ is a Lagrange multiplier. The resulting 
equation that connects $\lambda$, $x_A$, $x_B$ and $\gamma$ 
is ${\lambda (\lambda^2 - 4)} [\lambda + 2 (x_B - x_A)] 
= 2 \gamma$. For any $\gamma$, the above equation has 
three solutions, two of which are physically relevant. 
The one appears for $0 \le \lambda \le 2(x_A - x_B) 
= 2 (2 x_A - 1)$, which is $\le 2$, and the other 
one for $\lambda \ge 2$. The first solution gives 
the critical value of $\gamma$, $\gamma_{\rm cr}$, 
which gives a zero slope of the spectrum $E_0/N$ 
for $0 \le l \le 1$, at $l=x_A^-$ as function of 
$x_A$, namely 
\begin{eqnarray}
  \gamma_{\rm cr} = (3/2)/(4 x_A-3).
\label{dcr}
\end{eqnarray}
The above expression not only gives the exact value of 
$\gamma_{\rm cr}$ for $x_A = 1$ and $x_B = 0$ (which is
3/2, as mentioned earlier), but also for any (allowed) 
value of $x_A$. Since the above function diverges for 
$x_A \to 3/4$, persistent currents are only possible 
for $3/4 < x_A \le 1$. 

In the intervals of higher angular momentum, $n \le l \le 
n+1$ with $n \neq 0$, the situation with stability is rather 
different. According to Eq.\,(\ref{Bl}) the periodic part 
of the dispersion relation $e(l)$ repeats itself in each 
of these intervals with a slope that is equal to $(n+1)^2 
- n^2 = 2 n + 1 = 3, 5, 7, \dots$ For $n \neq 0$ one 
has to use the other solution for $\lambda > 2 (x_A - 
x_B)$. For the case of only one component, $x_A = 1$ 
and $x_B = 0$, this solution implies that persistent 
currents are stable for the values $\gamma_{\rm cr} = 
(2 n + 1) (2 n + 3)/2$, at $l = n+1$. While the above 
states support persistent currents, as soon as $x_B$ 
becomes nonzero -- even if $x_B \to 0$ but finite -- 
the other solution that lies in the interval $0 \le 
\lambda \le 2(x_A - x_B)$ has a lower energy, and 
destabilizes the current. In other words, the 
currents are very fragile with respect to admixtures 
of a second species of atoms. As a result, the system 
cannot support persistent currents at any interval 
other than the first one with $n \neq 0$, for $x_B 
\neq 0$. Figure 1 shows $\gamma_{\rm cr}$ of 
Eq.\,(\ref{dcr}), as well as the points corresponding
to $\gamma_{\rm cr} = (2 n + 1) (2 n + 3)/2$ for 
$n = 1, 2$, and 3.

To gain some physical insight on the above results, 
we note that for $0 \le l 
\le 1$, since the system is in the state $\phi_A = 
\Phi_1$ and $\phi_B = \Phi_0$ at $l=x_A$, it may 
reduce its angular momentum by either transferring 
some atoms of species $A$ from $\Phi_1$ to $\Phi_0$, 
or some atoms of species $B$ from $\Phi_0$ to 
$\Phi_{-1}$. However, the second option is 
energetically expensive because the angular 
momentum of $\Phi_{-1}$ is opposite to the angular 
momentum of the system. In the second interval
$1 \le l \le 2$ (and in any higher one) the system 
is in the state $\phi_A = \Phi_2$ and $\phi_B = 
\Phi_1$ when $l = 1 + x_A$. In this case, however, 
the most efficient way for the gas to reduce 
its angular momentum is to transfer atoms of 
species $B$ from $\Phi_1$ to $\Phi_0$, and not 
to transfer atoms of species $A$ from $\Phi_2$ 
to $\Phi_1$, as in the first interval. It is 
precisely this asymmetry between the first and 
any other interval that allows stable persistent
currents in the first interval only, but not in any
other.

\begin{figure}
\includegraphics[width=8.5cm,height=5.5cm]{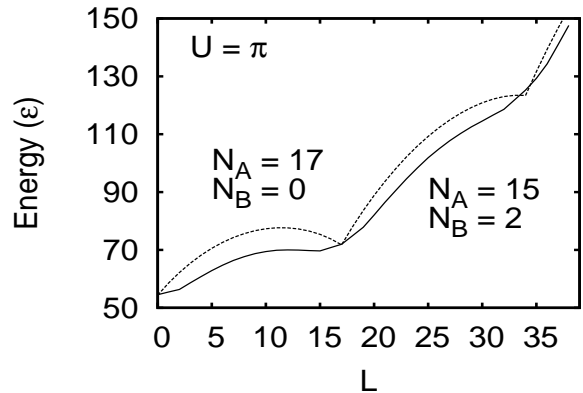}
\caption{The lowest eigenergies of the Hamiltonian
for $N_A = 17, N_B = 0$ (higher, dashed curve), as 
well as for $N_A=15, N_B = 2$ (lower, solid curve), 
with $U = \pi$, in the range $0 \le L \le 38$, in 
the truncated space of single-particle states with 
$|m| \le 7$.}
\label{FIG2}
\end{figure}

{\it Beyond the mean-field approximation.} 
To go beyond the mean-field approximation, we
have also performed numerical diagonalization of 
the Hamiltonian for fixed numbers of $N_A$, $N_B$
and $L$ units of angular momentum. In the case of
one component, we have confirmed the results
derived within the mean-field approximation 
$\gamma_{\rm cr} = 3/2$ for $n=0$, and 
$\gamma_{\rm cr} = 15/2$ for $n=1$. What is 
even more interesting is the lowest eigenenergy 
of the Hamiltonian for $N_A = 17$, $N_B = 0$, 
as well as for $N_A = 15$, $N_B = 2$, in the 
range $0 \le L \le 38$, incuding all the 
single-particle states with $|m| \le 7$, for 
$U = \pi$, which is shown in Fig.\,2, [the 
corresponding value of $\gamma$ has to be 
calculated according to the formula $\gamma 
= (N-1) U /(2 \pi)$, which gives $\gamma = 8$]. 
Figure 2 indicates clearly the metastability 
of the currents for $L = N_A$, and $L = 2 N_A$ 
when there is only one component. With the 
addition of even a small second component, the 
local minimum around $L = 2 N_A$ disappears, 
destroying the metastable current, while the 
minimum around $L = N_A$ still exists 
[$\gamma_{\rm cr} \approx 2.83$, according to 
Eq.\,(\ref{dcr})], in agreement with the 
mean-field approximation.

We have also found numerically that for $0 \le L 
\le N_B$ (and $N_A \le L \le N_A + N_B$), the 
(whole) excitation spectrum is given by the formula
$E_q(L) = L + U/(2 \pi) [ q^2 + (N + 1 - 2 L) q + 
 N (N - 1)/2 - L]$, where $q = 0, 1, 2, \dots$
in the truncated space of single-particle states 
with $m=0$ and 1 (the only ones which are 
macroscopically occupied in the limit of large $N$). 
The lowest energy per particle $E_0(L)/N = l + \gamma/2$ 
agrees with the result of mean-field in the limit 
$N \to \infty, L \to \infty$ with $L/N = l$ (finite), 
and $N U$ finite.

A more specific case of the above spectrum may even 
be derived analytically with use of the Bogoliubov 
transformation, for $L=N_B$ (or $L=N_A$), within the 
same truncated space of the single-particle states 
with $m=0$ and 1. Within the Bogoliubov approximation, 
the Hamiltonian takes the form in this case
\begin{eqnarray}
   H = N_B +  U /(2 \pi) \left[ N (N-1)/2 + \right.
\phantom{XXXXXXXXX}
\nonumber \\
\left.  + (N/2) (a_1^{\dagger} a_1
+ b_0^{\dagger} b_0) + \sqrt{N_A N_B} (a_1 b_0 + a_1^{\dagger}
b_0^{\dagger}) \right],
\end{eqnarray}
where $a_1$ is the annihilation operator of a boson 
of species $A$ with angular momentum $m=1$, and $b_0$ 
is the annihilation operator of species $B$ with
$m=0$. This Hamiltonian is diagonalized with a 
Bogoliubov transformation, which implies that the 
eigenvalues are (assuming, for example, that $N_A > N_B$),
\begin{eqnarray}
  {\cal E}_q (N_B)= N_B + 
\frac U {2 \pi} \left[ \frac N 2 (N-2) + (N_A - N_B) 
(2 q + 1) \right].
\end{eqnarray}
We then find that the difference $E_q(L=N_B) - 
{\cal E}_q(L=N_B) = U q(q+1) \propto 1/N$, 
and thus vanishes for large $N$. 

{\it Conclusions.} This study provides an
interesting illustration of the physical origin 
of persistent currents and, more generally, of 
superfluidity. The extra degrees of freedom due 
to the second component, combined with the 
assumed one-dimensionality and the periodicity
of the Hamiltonian, introduce novel physical 
effects, which have not been known in the physics 
of the ``traditional" superfluids. 

More specifically, 

(i): In one-component systems, sufficiently high 
values of the coupling give rise to persistent currents 
\cite{Leggett}. In the present case, unless the 
population of the second species is sufficiently small 
-- in which case one goes back to the one-component 
case -- the second species provides an 
energetically inexpensive way for the system to 
get rid of its circulation: the node that is 
necessary to form in the component that carries 
the circulation, in order for the circulation
to escape from the ring, is filled by the second 
component, very much like the coreless vortices
studied in higher dimensions. 

(ii) The reduced dimensionality introduces another 
remarkable effect: while metastability of persistent 
currents is absent in two-component systems in 
higher dimensions \cite{Ho,BCKR}, here the assumed 
one-dimensional motion makes it possible for 
persistent currents to be stable, at least under 
specific conditions.

(iii) The assumed periodicity in the Hamiltonian 
reflects itself on the dispersion relation, which is 
quasi-periodic, as in the one-component
problem. On the other hand, while persistent
currents corresponding to the first interval of the
angular momentum of the quasi-periodic part of the 
spectrum are stable, for higher values of the 
angular mometum, persistent currents are highly 
fragile, even for a very small admixture of a second 
species. This result is also in sharp contrast to the 
one-component case. 

The results presented in our study definitely deserve 
experimental investigation, in order for our predictions
to be confirmed. One effect that deserves both 
theoretical, as well as experimental attention is the 
deviation from the one-dimensional motion assumed here. 
One may argue that as this deviation increases, 
competing mechanisms change the behavior of the system, 
interpolating between one- and two- or three-dimensional 
motion, thus giving rise to rich physical effects. 

Last but not least, in addition to the above more 
theoretical remarks, the large degree of tunability 
of the persistent currents that we have demonstrated 
here, also makes these systems very appealing 
in terms of future technological applications.

\end{document}